\documentclass[article, notitlepage,superscriptaddress, longbibliography]{revtex4-2}
\usepackage{natbib}
\usepackage{graphicx}
\usepackage{amssymb}
\usepackage{amsmath}
\usepackage{ifthen}
\usepackage{braket}
\usepackage{xcolor}
\usepackage{bm}
\usepackage{bbm}
\usepackage{bbm}
\usepackage{cancel}
\usepackage{maybemath}
\usepackage{comment}
\usepackage{siunitx}
\usepackage{float}
\usepackage{dcolumn}
\newcolumntype{d}[1]{D{.}{.}{#1}}
\usepackage{subfigure}

\newcommand{\dd}{\mathrm{d}}
\newcommand{\ii}{\mathrm{i}}
\newcommand{\ee}{\mathrm{e}}

\def\dd{{\mathrm{d}}}
\def\ii{{\mathrm{i}}}
\def\ee{{\mathrm{e}}}

\newcommand{\calL}{\mathcal{L}}
\newcommand{\LASY}{\hat{\calL}_{\cancel{\rm SYM}}}

\newcommand{\plus}{{\mbox{\bf{\tiny +}}}}

\newcommand{\HH}{{\rm H}}
\newcommand{\HHbar}{\overline{\rm H}}

\definecolor{garrosgreen}{rgb}{0.1, 0.4, 0.1}
\definecolor{dartmouthgreen}{rgb}{0.05, 0.5, 0.06}
\definecolor{duelferred}{rgb}{0.7, 0.2, 0.1}
\definecolor{cambridgeblue}{rgb}{0.1, 0.3, 1.0}
\definecolor{oxfordblue}{rgb}{0.05, 0.2, 0.7}

\newcommand{\addrRolla}{Department of Physics and LAMOR,
Missouri University of Science and Technology,
Rolla, Missouri 65409, USA}

\newcommand{\addrDebrecen}{Hungarian Academy Institute
for Nuclear Physics (ATOMKI), Debrecen, Hungary}

\begin{document}

\title{Antimatter Free-Fall Experiments and Charge Asymmetry} 

\author{Ulrich David~Jentschura}
\affiliation{\addrRolla}
\affiliation{\addrDebrecen}

\begin{abstract}
We propose a method by which one could use modified
antimatter gravity experiments in order to perform a high-precision
test of antimatter charge neutrality. The proposal is based on the application
of a strong, external, vertically oriented electric field during an antimatter free-fall 
gravity experiment in the gravitational field of the Earth. The proposed 
experimental setup has the potential to drastically improve the limits on 
the charge-asymmetry parameter ${\overline \epsilon}_q$ of antimatter.
On the theoretical side, we analyze possibilities to describe a putative 
charge-asymmetry of matter and antimatter, proportional to the 
parameters $\epsilon_q$ and ${\overline \epsilon}_q$, by Lagrangian methods. 
We found that such an asymmetry could be described by
four-dimensional Lorentz-invariant operators
that break CPT without destroying the locality of the field theory.
The mechanism involves 
an interaction Lagrangian with field operators decomposed into
particle or antiparticle field contributions. Our Lagrangian is otherwise Lorentz, as
well as PT invariant. Constraints to be derived on the 
parameter ${\overline \epsilon}_q$ do not depend on the assumed 
theoretical model.
\end{abstract}

\maketitle

\section{Introduction}

The CPT theorem~\cite{Lu1954,Lu1957} encompasses large classes
of local field theories with Lorentz-invariant, local, and Hermitian Lagrangian terms.
All theories investigated in~\cite{Lu1954,Lu1957}
are invariant under the combined action of charge conjugation (C), parity inversion
(P), and time reversal (T). 
Due to the (almost) universal character of the 
theories studied in~\cite{Lu1954,Lu1957},
the accepted assumption is that CPT invariance 
in an inalienable symmetry of nature. 
Here, we explore the possibility that, 
by enlarging the class of the possible interactions, it
is possible to evade the CPT theorem by adding Hermitian, local, yet
CPT-violating interactions that lead to long-range,
gravity-mimicking, CPT-violating interactions, and~still conserve
Lorentz invariance.

Such interactions have 
been investigated in~\cite{HuDe1992,BrEtAl2011,AmEtAl2014,AhEtAl2016} 
with a separate charge asymmetry parameter for antimatter 
being introduced on an \emph{ad~hoc} basis.
We, here, explore if such asymmetry parameters
could be introduced via 
a separation of the quantum-field theoretical fermion operators into
positive-energy and negative-energy components,
and we discuss a possible related experimental \emph{ansatz}. 

Such interactions violate a
fundamental principle of the construction of quantum field 
theories, namely, gauge invariance. 
This is the primary reason why one can introduce 
putative charge-asymmetry of matter and antimatter 
by four-dimensional Lorentz-invariant operators that break CPT
without destroying the locality of the field theory.
Our interaction terms describe the exchange of virtual bosons but exclude virtual annihilation channels.
While, with~this approach, several difficulties associated with nonlocal 
and Lorentz-violating field theories can be avoided, 
gauge invariance is violated.

The experimental proposal discussed here is independent of the 
theoretical model introduced; the aim is to set limits for
the charge asymmetry parameter for antimatter,
which could otherwise 
be introduced on an ad~hoc basis~\cite{HuDe1992,AmEtAl2014,AhEtAl2016}.
Our approach is based on the 
explicit assumption that matter--antimatter
symmetry perfectly holds for the gravitational interaction.
There are strong theoretical arguments to support this
assumption~\cite{Po2010,Je2020physics}. This conjecture
also is compatible with the experimental results
reported in~\cite{AmEtAl2013mdpi}, which will hopefully 
be improved in the near future.

From a phenomenological point of view, the~
decisive characteristic of our models is the
fact that they allow for the presence of a different
charge asymmetry parameter for matter versus antimatter.
Such effects have been discussed in the literature,
independent of a quantum-field theoretical
underlying formulation~\cite{BrEtAl2011,AmEtAl2014,AhEtAl2016}. 
Our proposal for an improvement of the 
charge asymmetry parameter for antimatter 
is connected with antimatter gravity
experiments with~the aim of comparing a residual 
electrostatic force on an antihydrogen atom
with the gravitational force. 

Tests of the charge neutrality of 
matter (and antimatter) have recently attracted considerable
attention~\cite{BrEtAl2011,AmEtAl2014,AhEtAl2016}.  A~conceivable
electric charge of the neutron was investigated in~\cite{ShEs1956}. 
An~excellent review on various theoretical models
allowing for charge asymmetry is presented in~\cite{FoEtAl1990}.
An essential assumption underlying our investigations
is that the gravitational interaction for antimatter
in the gravitational field of the Earth
is the same as for matter.
There are strong theoretical arguments to 
support this initial assumption~\cite{Po2010,JeNo2013pra,Je2020physics}.

Past efforts~\cite{ScBa1966,WiFa1967,WiFa1968}
to measure the gravitational interaction 
of electrons and positrons have suffered
from the problem of eliminating electric-field 
effects. This has been a tremendous
problem that has never been solved convincingly in
experiments~\cite{WiFa1967,WiFa1968}.
It is connected with the~elimination of
the influence of the electric field
generated by the electrons in the metal or other
material (drift tube) surrounding the 
fall line of the charged leptons.

Theoretical arguments~\cite{ScBa1966} suggested that, 
under the presence of the electric field
generated by the electrons, the~total gravitational
acceleration of electrons converges to zero,
while that of positrons would be twice the
acceleration due to gravity.
Indeed, experiments with positrons were not succesfully reported
despite considerable invested effort~\cite{ScBa1966,WiFa1967,WiFa1968}.
We argue the other way and show that,
in a gravitational experiment carried out with 
(supposedly) electrically neutral particles, the~influence
of any deliberately introduced, strong, external,
electric field allows one to display the 
effect of a residual charge excess,
leading to a test of charge~neutrality.

The influence of a hypothetical charge excess
in antimatter on the dynamics can be 
explored effectively in a
strong, uniform, vertical external electric
field. This is because any potential charge
asymmetry is  compared to the extremely 
weak gravitational force. 
Our estimates, reported in this article,
suggest that limits on the charge asymmetry
of antimatter could be improved by many orders 
of magnitude if an antihydrogen gravity experiment
is done in the presence of a strong external
electric~field.
SI mksA units are used in this article 
unless stated otherwise.

%
%
\section{Charge Symmetry and~Gravity}
\label{sec3}

Let us assume that electron and proton charges do not
quite add up to zero so that there is an ever so slight
residual charge to be associated with a
hydrogen atom, an~idea originally 
formulated by Einstein at the 1924 Lucerne Meeting of the
Swiss Physical Society as~was explicitly
mentioned in~\cite{PiKe1925,PiKe1925cite}.
In accordance with~\cite{BrEtAl2011},
we parameterize a putative charge excess
as follows,
\begin{align}
q_e =& \; -|e| \,, \qquad q_p = |e| \, (1 + \epsilon_{p-e}) \,, \\
\epsilon_{p-e} =& \; \frac{q_e + q_p}{|e|} \,, \qquad
\epsilon_n = \frac{q_n}{|e|} \,.
\end{align}
Here, $q_e = e$ is the electron charge, and
$|e|$ is its modulus, while $q_p$ and $q_n$ are the proton 
and neutron charges, respectively.
If we assume, with~\cite{BrEtAl2011},
charge conservation in the $\beta$ decay of the 
neutron, then the charge-neutrality violating
parameter $\epsilon_n$ for the neutron 
{ (let us be clear that 
the neutron acquires an 
infinitesimal electric charge under the assumptions made in 
~\cite{BrEtAl2011})} becomes
\begin{equation}
\label{assump1}
\epsilon_n = \epsilon_{p-e} \equiv \epsilon_q \,.
\end{equation}
If a body containing $Z$ protons and electrons,
as well as $N$ neutrons is measured as being neutral
with sensitivity $\delta q$, the~one can
obtain a limit on $|\epsilon_q| $, which is on the 
order of 
\begin{subequations}
\begin{align}
\left| Z \epsilon_{p-e} + N \epsilon_n \right| = & \;
(Z + N) |\epsilon_q| \leq \delta q \,,
\\[0.1133ex]
|\epsilon_q| \leq & \; \frac{\delta q}{(Z+N) |e|} \,.
\end{align}
\end{subequations}
The essential idea of the acoustic method used in~\cite{BrEtAl2011}
is that, under~the assumption of a nonvanishing charge
asymmetry in matter, electromagnetic waves incident on an
electrically neutral gas would set the gas atoms in 
motion, inducing sound waves. However, this method
of determining limits on $| \epsilon_q |$ is not free
from pitfalls and requires a considerable additional mathematical 
formalism in the evaluation of the experiment. 
For example, according to a note to Table~I of~\cite{StMoTr1967},
data published in the previous work~\cite{Ki1960}
may exhibit inconsistencies.

A paper~\cite{DyKi1972} that initially claimed an accuracy on the level of $10^{-23}$ 
for $|\epsilon_q|$
has recently been questioned in~\cite{BrEtAl2011},
with the claim that their result on $|\epsilon_q|$ could
not be considered to be better than $10^{-19}$ 
if all inaccuracies and neglected systematic 
effects in the paper~\cite{DyKi1972}
are properly taken into account. 
The paper~\cite{BrEtAl2011} also indicates  additional 
rectification of the analysis of the resonant modes in the 
gas-filled capacitor used in the previous experiment~\cite{DyKi1972}.
Table I of~\cite{BrEtAl2011} contains a comprehensive
compilation of previous measurements of $\epsilon_q$. We will use
their result, given in an unnumbered equation on the second-to-last
page of~\cite{BrEtAl2011},
\begin{equation}
\label{eps_q}
\epsilon_q = (-0.1 \pm 1.1) \times 10^{-21} \,,
\end{equation}
for matter particles (both a hydrogen atom as well as the 
constituent atoms of the Earth).
Limits on the charge asymmetry of matter have also been
derived on the basis of model-dependent 
astrophysical methods~\cite{Se2000,CaFe2000}.
Separate investigations  put limits on the neutrality of neutrinos
on the basis of 
astrophysical observations~\cite{BeRuFe1963,BaCo1987,Si1989,SePa1996,Ra1999}.

In contrast, the~constraints on
charge neutrality for antimatter are looser by many orders of magnitude. Tests
on the electric charges of positrons and antiprotons 
can be derived from measurements of their cyclotron
resonance frequencies and from spectroscopic data~\cite{HuDe1992}.
The most recent direct tests~\cite{AmEtAl2014,AhEtAl2016} 
revealed a $1\sigma$ limit $68.3$\% confidence level)
\begin{equation}
\label{eps_qbar}
|{\overline \epsilon}_q| \leq 7.1 \times 10^{-10}  
\end{equation}
for antimatter. Here, the~parameters for antiparticles are given by
\begin{align}
q_{\overline e} =& \; |e| \,, \qquad 
q_{\overline p} = -|e| \, (1 + 
\epsilon_{{\overline p}-{\overline e}}) \,, \\
\epsilon_{{\overline p}-{\overline e}} =& \; \frac{q_{\overline e} + 
q_{\overline p}}{-|e|} \,, \qquad
\epsilon_{\overline n} = \frac{q_{\overline n}}{-|e|} \,.
\end{align}
The parameters ${\overline e}$, ${\overline p}$, and 
${\overline n}$ stand for the positron, antiproton, and~
antineutron, respectively.
We shall also make the assumption that charge 
is conserved in the $\beta$ decay of the antineutron and 
write
\begin{equation}
\label{assump2}
\epsilon_{\overline n} = 
\epsilon_{{\overline p}-{\overline e}} \equiv {\overline \epsilon}_q \,.
\end{equation}
Furthermore, we shall assume, as~demonstrated in 
a number of experiments~\cite{WeStCo1979,PeChCh1999,SaEtAl2019},
that gravitational and gravity-like interactions 
are equivalently realized on the microscopic (atomic)
and macroscopic~level.

The electric charge asymmetry
of antihydrogen, if~it exists, is not
necessarily opposite to that found in hydrogen.
However, there might be good arguments
to support this conjecture.
Namely, electrons and positrons
constitute a particle--antiparticle pair
and, therefore, are described by the same
Dirac equation, which predicts that the
electric charges of the positron and electron
add up to zero.
The same applies to protons and antiprotons.
However, one observes that leptons and hadrons
are still two completely different particle species.
Therefore, it could appear easier to speculate about the
broken electric neutrality of a
hydrogen atom rather than the 
broken electric neutrality of, say, positronium.

\section{Charge Asymmetry and CPT~Violation}

Typically (see~\cite{PiKe1925,PiKe1925cite,BrEtAl2011}),
the charge-asymmetry parameters $\epsilon_q$ and ${\overline \epsilon}_q$ 
are used as \emph{ad~hoc} parameters without~
any attempt being made to formulate an underlying 
quantum field theory that could describe the charge-symmetry violation.
Let us attempt to realize somewhat higher ambitions and explore 
candidate models. In~the field-theoretical sense,
let us explore if
a slight breaking of the charge symmetry 
could be formulated in terms of 
a gauge-symmetry breaking (GB)
modification of the quantum
electrodynamic (QED) interaction
Lagrangian. Denoting field operators by a hat,
we write the Lagrangian
(we temporarily switch to the natural unit system 
with $\hbar = c = \epsilon_0 = 1$, as is customary
in particle physics)
\begin{align}
\label{LGB}
\hat\calL_{\rm GB} = & \;
\sum_{f=e,p,n} \hat{\overline\psi}_f
(\ii \gamma^\mu \, \partial_\mu - m) \,
\hat{\psi}_f - \frac14 F_{\mu\nu} \, F^{\mu\nu}
\nonumber\\[0.1133ex]
& \; -e \, 
\left( \hat{\overline \psi}_e \, \gamma^\mu \, \hat{\psi}_e 
- \hat{\overline \psi}_p \, \gamma^\mu \, \hat{\psi}_p \right) \, {\hat A}_\mu
\nonumber\\[0.1133ex]
& \; - \epsilon_q \, |e| \,
\left(\hat{\overline \psi}_p \, \gamma^\mu \, \hat{\psi}_p 
+ \hat{\overline \psi}_n \, \gamma^\mu \, \hat{\psi}_n \right) \, {\hat A}_\mu \,.
\end{align}
Here, the~electron-positron field operator is $\psi_e$,
while the composite spin-1/2 operator for the 
proton--antiproton field is $\psi_p$, and~the 
neutron--antineutron field could be described 
by a spin-1 generalization of the Dirac equation (see~\cite{Mo2010}).
The quantized photon field is described by the 
four-vector field operator ${\hat A}_\mu$,
which enters the field-strength tensor operator
${\hat F}_{\mu\nu} = \partial_\mu {\hat A}_\nu - \partial_\nu {\hat A}_\mu$,
where $\partial_\mu \equiv \partial/\partial x^\mu$ is the 
partial derivative with respect to the space-time coordinate $x^\mu$.
In the form of $\hat\calL_{\rm GB} $ given
in Equation~\eqref{LGB},
we use the assumption~\eqref{assump2}.
As a side remark, we note that the 
current conservation implies that 
$\epsilon_{p-e} = 2 \epsilon_u + \epsilon_d $,
where the valence quark couplings for the 
up and the down quark are $\epsilon_u$ and $\epsilon_d$, 
respectively. 

From the form of the Lagrangian~\eqref{LGB},
it follows that electrons (and consequently positrons) 
carry a charge $\pm e$,
while protons (and antiprotons) carry a charge 
$\pm (1+\epsilon_{p-e}) \, |e|$.
This results in a hydrogen atom having a charge 
$\epsilon_{p-e} \, |e|$,
while antihydrogen atoms carry a charge
$(-\epsilon_{p-e}\,|e|)$.
One might, therefore, assume that
$\epsilon_q = -{\overline \epsilon}_q$.

As there exist very stringent limits on 
$\epsilon_q$ (for particles, see~\cite{BrEtAl2011}),
it is  interesting to explore the possibility of 
different charge-asymmetry parameters for particles
and antiparticles, i.e.,
a Lagrangian with two different 
parameters $\epsilon_q$ and ${\overline \epsilon}_q$. 
One may, thus, explore the phenomenological 
consequences of the Lagrangian
\begin{align}
\label{LASY}
\LASY= & \;
\sum_{f=e,p,n} \hat{\overline\psi}_f
\left(\ii \gamma_f^\mu \; \partial_\mu - m\right) \,
\hat{\psi}_f - \frac14 {\hat F}_{\mu\nu} \, {\hat F}^{\mu\nu}
\nonumber\\[0.1133ex]
& \; -e \,
\left( \hat{\overline \psi}_e \, \gamma^\mu_e \, \hat{\psi}_e
- \hat{\overline \psi}_p \, \gamma^\mu_e \, \hat{\psi}_p \right) \, {\hat A}_\mu
\nonumber\\[0.1133ex]
& \; - \epsilon_q \, |e| \,
\left(\hat{\overline \psi}^{(+)}_p \, \gamma^\mu_p \, \hat{\psi}^{(+)}_p
+ \hat{\overline \psi}^{(+)}_n \, \gamma^\mu_p \, \hat{\psi}^{(+)}_n \right) \, 
{\hat A}_\mu 
\nonumber\\[0.1133ex]
& \; + {\overline \epsilon}_q \, |e| \,
\left(\hat{\overline \psi}^{(-)}_p \, \gamma^\mu_n \, \hat{\psi}^{(-)}_p
+ \hat{\overline \psi}^{(-)}_n \, \gamma^\mu_n \, \hat{\psi}^{(-)}_n \right) \, 
{\hat A}_\mu \,.
\end{align}
Here, we have induced CPT violation by decomposing a general fermionic
field operator $\hat\psi(x)$ into a positive-frequency
matter contribution $\psi^{(+)}(x)$ and a negative-frequency
antimatter contribution $\psi^{(-)}(x)$, as follows,
%
%
\begin{align}
\label{decomp}
\hat\psi(x) =&\; \psi^{(-)}(x) + \psi^{(+)}(x) \,,
\\[0.1133ex]
\hat\psi^{(+)}(x) = & \; \sum_s \int \frac{\dd^3 p}{(2 \pi)^3} \, \frac{m}{E} \,
a_s(\vec p)\, u_s(\vec p) \, \ee^{-\ii p \cdot x} \,,
\\[0.1133ex]
\hat\psi^{(-)}(x) = & \; \sum_s \int \frac{\dd^3 p}{(2 \pi)^3} \, \frac{m}{E} \,
b^\plus_s(\vec p) \,v_s(\vec p) \,\ee^{\ii p \cdot x} \,.
\end{align}
In view of the impossibility to transform positive-energy states
into negative-energy states via Lorentz transformations (due the mass gap
for Dirac particles), this decomposition is Lorentz invariant (for details,
see~\cite{JeNaTr2021plb}).
It is useful to remark that the decomposition is also  PT invariant.
The decomposition of the effective proton and neutron 
field operators into positive-frequency and negative-frequency
contributions has led to CPT violation and allows for different
charge-asymmetry parameters $\epsilon_q$ (for particles,
i.e., for~hydrogen) and ${\overline \epsilon}_q$ (for antiparticles,
i.e., for~antihydrogen).

A closer inspection reveals that the CPT-violating 
terms in Equation~\eqref{LASY} (those proportional to 
$\epsilon_q$ and ${\overline\epsilon}_q$) 
only modify the exchange of virtual photons between 
protons (and neutrons, due to the infinitesimal neutron charge); 
however, they do not lead to any annihilation of a proton--antiproton 
pair into a virtual photon. 
In that sense, the~physics deduced from the interaction~\eqref{LASY}
implies a slight modification of the electromagnetic interaction 
between protons as compared to electrons, affecting the 
charge neutrality of the hydrogen and antihydrogen atoms.

This leads to a violation of the electromagnetic gauge 
invariance (both because of the slightly different electron
and proton charge and~because of the separation of the 
field operator in the interaction Hamiltonian
into positive-energy and negative-energy components).
The consequences of this assumption of astrophysical
scales need to be explored.
Two aspects can be mentioned here. 

First, 
let us remember that a conceivable charge asymmetry of matter 
has been discussed as a possible (partial) explanation for the expansion
of the Universe~\cite{Hu1957,LyBo1959}.
Even if the commonly accepted explanation involves a
nonvanishing cosmological constant~\cite{BaEtAl1999mnras,Pl2016},
it would be interesting to explore conceivable additional 
contributions to the expansion of the Universe due to 
antimatter and matter charge asymmetries.

There is a second aspect that might be even more interesting.
Namely, a~closer inspection reveals that 
the infinitesimal charge excess of the proton 
against the electron mimics, in~the nonrelativistic 
limit, a~gravitational interaction
(see also Equation~\eqref{VVV} below).
However, the~relativistic corrections to the 
gravitational interaction are known to 
be different for gravity as compared to 
electromagnetism~\cite{BjDr1964,JeNo2013pra}.

If a nonvanishing charge asymmetry parameter were found
for antimatter or matter,
then one might need to investigate the effective
modifications of the gravitational interaction 
(now adjusted for the retardation corrections
to the electromagnetic admixtures due to the 
charge asymmetry). These could potentially have interesting 
connections to the observed discrepancies between 
astrophysical observations and the accepted theory 
of gravitation and, thus, to~the dark matter problem.

The free photon field term in the Lagrangian density~\eqref{LASY}
involves the field-strength tensor 
\begin{equation}
{\hat F}_{\mu\nu} = \partial_\mu {\hat A}_\nu - \partial_\nu {\hat A}_\mu \,.
\end{equation}
A closer inspection reveals that, 
in the nonrelativistic limit, 
the Lagrangian~\eqref{LASY} leads to an effective interaction
between hydrogen (H) and antihydrogen atoms ($\HHbar$) of
(we now switch back to SI mksA units for the 
remainder of this article)
\begin{subequations}
\label{VVV}
\begin{align}
V_{\HH \HH}(R) =& \;
\frac{\epsilon_q^2 \, e^2}{4 \pi \epsilon_0 \, R} \,,
\\[0.1133ex]
V_{\HHbar\HH}(R) 
=& \; \frac{\epsilon_q \, {\overline \epsilon}_q \, e^2}{4 \pi \epsilon_0 \, R} \,,
\\[0.1133ex]
V_{\HHbar\HHbar}(R) 
=& \; \frac{{\overline \epsilon}_q^2 \, e^2}{4 \pi \epsilon_0 \, R} \,,
\end{align}
\end{subequations}
where $R$ is the interatomic distance.
None of the interaction Lagrangians 
investigated in~\cite{Lu1954,Lu1957}
involve the splitting of a field operator into
positive-frequency and negative-frequency parts. 
Our explicit separation of the field operators
into positive- and negative-energy components
in Equation~\eqref{LASY} offers the possibility of introducing
different charge asymmetry parameters for 
matter and~antimatter.

\section{Possible Implications on Charge Asymmetry of~Antimatter}

Based on the Lagrangian~\eqref{LASY},
our goal was to investigate whether or
not the limit given in Equation~\eqref{eps_qbar},
namely, $|{\overline \epsilon}_q| \leq 7.1 \times 10^{-10} $
could be improved by a simple experimental 
arrangement: Our suggestion was to place 
freely falling antihydrogen atoms into
a uniform, vertically oriented, electric field,
which would enable us to compare the
gravitational force acting on the (supposedly)
neutral antihydrogen atom to~the 
gravitational force. We would, thus, intend to make the 
background electric field large, not small
(as was otherwise attempted in 
~\cite{ScBa1966,WiFa1967,WiFa1968}).

\begin{figure}[H]

\begin{center}
\begin{minipage}{0.71\linewidth}
\begin{center}
\includegraphics[width=0.42\linewidth]{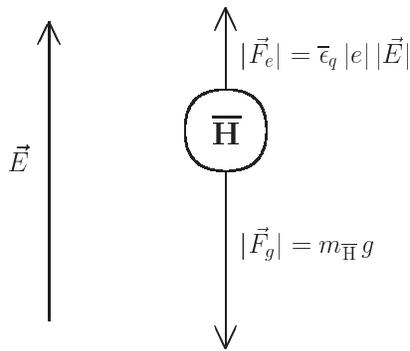}
\end{center}
\caption{\label{fig1} The schematic of the proposed experiment involves
a freely falling antihydrogen atom ${\overline{H}}$ in a 
strong external, static, vertically oriented  electric field $\vec E$.
If the antihydrogen atom fails to be electrically neutral, then 
the gravitational force of magnitude 
$|\vec F_g| = m_{\overline H} \, g$
competes with the residual electrostatic force
$|\vec F_e| = {\overline \epsilon}_q \, |e| \, |\vec E|$, thus,
leading to a corresponding reduction of the resultant
force (or, to~an enhancement, upon~flipping the 
sign of the electric field).  In~proposing the experiment,
we assumed that the gravitational interaction for antimatter
in the gravitational field of the Earth
is the same as for matter~\cite{Po2010,JeNo2013pra,Je2020physics}.}

\end{minipage}
\end{center}

\end{figure}

An estimate of the achievable accuracy of 
$|{\overline \epsilon}_q|$ can be obtained 
by a simple calculation (see also Figure~\ref{fig1}).
The magnitude of the gravitational force on an antihydrogen atom is 
\begin{equation}
| \vec F_g | = m_{\overline{\rm H}} \, g \,,
\end{equation}
where $m_{\overline H}$ is the antihydrogen atom's mass.
The magnitude of the residual electric force on the antihydrogen atom is 
\begin{equation}
| \vec F_e | = {\overline\epsilon}_q \, |e| \, |\vec E| \,,
\end{equation}
where $|\vec E|$ is the (possibly strong) external, 
vertically oriented electric field.
Let us assume that we can experimentally establish that 
\begin{equation}
| \vec F_e | < \chi \, | \vec F_g |  \,,
\end{equation}
i.e., that the residual putative electric force
is less than a fraction $\chi$ of the gravitational force.
Let us also parameterize the 
field strength of the vertically oriented field 
as $|\vec E| = |\vec E|_{\rm SI}\, \frac{{\rm V}}{\rm m}$,
where $|\vec E|_{\rm SI}$ is the magnitude of the field,
measured in volts per meter.
This sets a limit on $|{\overline \epsilon}_q|$,
which is of the order of 
\begin{equation}
\label{bound1}
{\overline\epsilon}_q <
\frac{\chi \, m_{\HHbar} \, g}{|e| \, |\vec E|}
= 1.02 \times 10^{-7} \frac{\chi}{ |\vec E|_{\rm SI}} \,.
\end{equation}
Typical Cockroft--Walton voltage
multipliers operate at voltages of 
of $2 \times 10^4 \, {\rm V}$ or 
more~\cite{CoWa1932a,CoWa1932b,We1969},
and recent developments easily reach the range 
or $10^5 \, {\rm V}$ (see~\cite{MaEtAl2016}).
Powerful tandem accelerators are known to operate
at $2.5 \times 10^7 \, {\rm V}$
(see~\cite{WiEtAl2009,GrEtAl2012}).
In view of Paschen's law~\cite{Pa1889}, the~
electric breakdown strength inside an antihydrogen trap
is of no concern in a trap held at a good vacuum
(below $10^{-7}$ atmospheric pressure). 
Here, we estimate, somewhat conservatively, that an 
electric field strength of 
$|\vec E| = 10^6 \, \frac{{\rm V}}{\rm m}$
can be realized in a dedicated experiment, 
corresponding to a value of $|\vec E|_{\rm SI} = 10^6$.

Under the further conservative assumption that the 
experiment establishes that the magnitude residual electric 
force is less than 10\% of the gravitational force ($\chi = 0.1$),
one could improve the limit on 
${\overline\epsilon}_q$ into the range
\begin{equation}
\label{bound2}
|{\overline\epsilon}_q | \lesssim 10^{-14} \,,
\end{equation}
potentially improving the limit~\eqref{eps_qbar} by many orders of magnitude,
leading to a drastic improvement of the charge neutrality parameter
for antimatter. Possible limitations due to 
systematic effects encountered in the experiment
(interatomic interactions and the spin coupling to the 
electric field) are discussed in the Appendix~\ref{appa}.

\unskip

%
%
\section{Conclusions}
\label{sec4}

In this work, we discussed possible antimatter charge asymmetry
characterized by an antimatter charge
asymmetry parameter ${\overline \epsilon}_q$.
Our considerations were motivated by the fact that 
the charge-asymmetry parameter ${\overline \epsilon}_q$ for 
antimatter might be different from the charge asymmetry 
parameter $\epsilon_q$ for particles (see Equation~\eqref{LASY}).
We, thus, advocate an electrically counterbalanced antimatter gravity 
for the potential determination of an improved bound on the 
charge-asymmetry parameter ${\overline \epsilon}_q$ 
for antimatter. 

In~the proposed experimental setup, a~strong,
vertically oriented, electric field is placed in the 
line of free fall of an antihydrogen atom. 
Any putative charge asymmetry of antimatter,
under the influence of the additional external field, would 
modify the resultant acceleration of the antihydrogen atom, 
which, in~the absence of charge asymmetry, would exclusively 
be due to gravity. 
Even under pessimistic parametric estimates, the
bounds on ${\overline\epsilon}_q$
could potentially be improved by many orders 
of magnitude when compared to the current limits
(see Equations~\eqref{eps_qbar},~\eqref{bound2} 
and~\cite{AmEtAl2014,AhEtAl2016}).


\acknowledgments{The authors acknowledge 
insightful conversations with Professor Michael Schulz.
Helpful conversations with Professors Zolt\'{a}n 
Trocs\'{a}nyi and Istv\'{a}n N\'{a}ndori also are gratefully acknowledged.
Support from the
National Science Foundation (Grants PHY--1710856 and PHY--2110294)
is also gratefully~acknowledged.}

%
%
\section{Interatomic Interactions and Spin-Orbit~Force}
\label{appa}

A potential issue with any experiment is
systematic effects. While details of 
the realization of the current proposal are beyond the scope of the 
current paper, some remarks on the role of 
non-gravitational matter--antimatter interactions and 
spin couplings to the electric field are in~order.

The nonretarded van-der-Waals interaction energy 
among hydrogen atoms,
in the leading approximation, is~\cite{Ko1967,DeYo1973,AdEtAl2017vdWi}
\begin{equation}
E_{\HH \HH}(R) \approx - \frac{D_6}{R^6} \,,
\qquad
D_6 = 6.499\, E_h\, a_0^6 \,,
\qquad
R \ll \frac{a_0}{\alpha} \,,
\end{equation}
where $E_h$ is the Hartree energy and $a_0$ is the Bohr radius
(the fine-structure constant is $\alpha$).
In the long-range limit, the~result changes due to 
retardation to~\cite{AdEtAl2017vdWi}
\begin{equation}
E_{\HH \HH}(R) \approx \; -\frac{23}{4 \pi}
\frac{\hbar c}{(4\pi\epsilon_0)^2}\frac{1}{R^7} \;
\alpha_{1S}(0) \, \alpha_{1S}(0) \,,
\qquad
R \gg \frac{a_0}{\alpha} \,,
\end{equation}
where
\begin{equation}
\alpha_{1S}\left(0\right) = \frac92 \frac{e^2 a_0^2}{E_h} 
\end{equation}
is the static polarizability of hydrogen.
In the long-range limit, the~interatomic interaction
is also referred to as the Casimir--Polder 
interaction~\cite{CaPo1948}.
Due to the $R^{-6}\ldots R^{-7}$ dependence,
the van-der-Waals (viz., Casimir--Polder) 
interaction can typically be considered as
negligible at large interatomic
separations. However, we  remember that, here, we compare it 
to gravitational interactions, and~thus a~closer look 
is~required.

Of interest in our context are matter--antimatter
interatomic interactions, i.e.,~the van-der-Waals 
and Casimir--Polder
interactions between hydrogen and antiydrogen atoms.
A closer inspection reveals that the both the 
van-der-Waals as well as the 
Casimir--Polder interaction between hydrogen and antihydrogen has
exactly the same coefficients and the same attractive
sign when~compared to the interaction between two hydrogen atoms,
i.e., one has 
\begin{equation}
\label{EHHeqEHH}
E_{\HHbar \HH}(R) = E_{\HH \HH}(R)  \,.
\end{equation}
This conclusion is not completely trivial.  The~(first-order) 
van-der-Waals Hamiltonian, given in Equation~(2c) of~\cite{JeEtAl2017vdWii},
actually changes sign in the transition from the 
hydrogen--hydrogen to the hydrogen--antihydrogen system.
This is evident because the nonretarded van-der-Waals 
Hamiltonian counts the electrostatic interactions between
the constituent particles of both atoms.
For the hydrogen--hydrogen system, one exemplary term in the 
van-der-Waals Hamiltonian is due to the interaction of 
the orbiting particle of atom $A$ (the electron)
with the nucleus of atom $B$ (the proton).

For the hydrogen--antihydrogen system, the~same term is 
replaced by the interaction of
the orbiting particle of atom $A$ (the electron)
with the nucleus of atom $B$ (the antiproton).
An obvious generalization of this consideration 
explains the overall sign change of the 
van-der-Waals Hamiltonian. However, the~$1/R^6$ 
van-der-Waals interaction is 
given by a {\em second}-order perturbation theory result
(see~\cite{JeEtAl2017vdWii,JeEtAl2019jpb1,MaEtAl2019jpb2}).
Thus, the~sign change of the van-der-Waals Hamiltonian leads to 
an invariant expression for the interaction energy,
in second-order perturbation theory.
Analogous considerations apply in the long-range limit,
where the virtual photon exchange between the two atoms
has to be formulated with the full energy dependence 
of the photon propagator, and~two mutually compensating
sign changes occur~\cite{CrTh1984}.
This justifies Equation~\eqref{EHHeqEHH}.

To analyze the role of interatomic 
interactions in our proposed test of antimatter charge 
neutrality, one should compare the interatomic 
interactions with the gravitational force acting 
on antihydrogen in the gravitational field of the Earth.
This calculation is easily done and reveals that the 
gravitational force on the Earth's surface dominates over the 
van-der-Waals interaction between two hydrogen or
antihydrogen atoms for all relevant interatomic distances
greater than a Bohr radius.
This observation, together with the functional form 
of the interatomic interactions, clarifies that 
interatomic interactions
do not affect the viability of our~proposal.

Another consideration is the role of the spin coupling 
of the electric field, especially in view of the 
fact that antihydrogen (just like hydrogen) has a 
magnetic moment largely determined by the positron spin
orientation (the antiproton spin contributes only 
a fraction to the total magnetic moment of the 
antihydrogen ground state).
Oscillating electric fields
can induce the spin flip of an electron~\cite{NoEtAl2007},
a fact that is also used in spintronics
(for a review, see~\cite{HiEtAl2020}).
The dominant term 
responsible for the spin coupling to the electric field is given 
by the spin-orbit (SO) coupling Hamiltonian~\cite{BjDr1964,ItZu1980},
which, in~SI mksA units, takes the form
\begin{equation}
\label{HSO}
H_{\rm SO} = -\frac{\hbar \,q}{4 m^2 c^2} \, 
\vec\sigma \cdot (\vec E \times \vec p) \,.
\end{equation}
Here, $q$ is the charge of the particle, $m$ is its mass,
$\vec\sigma$ denotes the vector of the Pauli spin matrices,
$\vec E$ is the electric field, 
and $\vec p$ is the particle momentum operator.
For a bound electron in a hydrogen-like ion of 
charge number $Z$, one replaces
$q \, \vec E \to - Z \, e^2 \vec r/(4 \pi \epsilon_0 r^3)$.
In this case, the~spin-orbit coupling reduces to the 
familiar Russell--Saunders coupling Hamiltonian, 
and $H_{\rm SO}$ is replaced as follows,
\begin{equation}
H_{\rm SO} \to 
\frac{\hbar^2 \, Z \alpha \, \vec\sigma \cdot \vec L}{4 m^2 c \, r^3} \,.
\end{equation}
As is well known~\cite{BeSa1957,ItZu1980}, the~Russell--Saunders
coupling determines the fine-structure of hydrogen in 
leading order. For~our proposed experiment, the
dominant electric field in Equation~\eqref{HSO} is the external, strong, vertical 
electric field, and~$\vec p$ is the electron 
momentum operator. The~momenta of the positron and of the 
antiproton in the freely falling antihydrogen atom 
can be transformed as follows into center-of-mass coordinates,
\begin{equation}
p_{e^+} = \frac{m_{e^+}}{m_{\HHbar}} \, \vec P + \vec p_r \,,
\qquad
p_{\overline p} = \frac{m_{\overline p}}{m_{\HHbar}} \, \vec P - \vec p_r \,,
\qquad
m_{\HHbar} = m_{e^+} + m_{\overline p} \,.
\end{equation}
Here, $\vec p_r$ is the relative momentum, which, 
in the ground state of antihydrogen, has the expectation value 
$\langle \vec p_r \rangle = \vec 0$.
We denote the positron mass as $m_{e^+}$, the antiproton mass as $m_{\overline p}$, 
and the total momentum of the antihydrogen atom by $\vec P$.
The appropriate spin-orbit coupling Hamiltonian for 
the coupling of the positron to the vertical electric 
field~is, therefore,
\begin{equation}
\label{HSOexp}
H_{\rm SO} = -\frac{\hbar |e|}{4 m_{e^+} m_{\HHbar} c^2} \, 
\vec\sigma_{e^+} \cdot (\vec E \times \vec P) \,,
\end{equation}
where $\vec\sigma_{e^+}$ is the vector of the positron spin matrices.
The classical trajectory of a freely falling 
body in the gravitational field of the Earth implies the 
relation
\begin{equation}
\vec P 
= -m_{\HHbar} \, g \, t \, {\hat{\rm e}}_z
= -m_{\HHbar} \, \sqrt{ 2 \, g \, h } \, {\hat{\rm e}}_z \,,
\end{equation}
where $g$ is the acceleration due to gravity, 
$t$ is the time, and~$h$ is the (downward) distance traveled 
by the antihydrogen atom in the gravitational field of the Earth.
Furthermore, ${\hat{\rm e}}_z$ is the unit vector in the 
$z$ direction. If~the external electric field is not 
perfectly aligned with the vertical, i.e.,
$\vec E \times \vec P \neq \vec 0$, 
then the spin-orbit energy has to be added to the 
kinetic energy acquired by the antihydrogen atom 
in free fall. Depending on the spin orientation of the 
positron, one has
\begin{equation}
\label{ESO}
E_{\rm SO} = \langle H_{\rm SO} \rangle 
= \pm \frac{\hbar |e| |\vec E| |\vec P| \, \sin\theta} {4 m_{e^+} m_{\HHbar} c^2} \,
= \pm \frac{\hbar |e| |\vec E| \sqrt{ g \, h } 
\, \sin\theta} {2 \sqrt{2} m_{e^+} c^2} \,,
\end{equation}
where $\theta = \angle( \vec E, \vec P)$.
We chose the quantization axis of the positron 
spin to be parallel to the direction of $\vec E \times \vec P$.

A force on the freely falling antihydrogen atom 
is generated when the spin-orbit energy $E_{\rm SO}$ 
becomes position-dependent, i.e.,~when 
there is a field gradient of the electric field $\vec E$.
Let us estimate that 
\begin{equation}
\left| \frac{\partial}{\partial x^i} \, E^j \right| 
\sim
\frac{ |\vec E| }{ L } \,,
\qquad 
i,j = \{ 1,2,3 \} \,,
\end{equation}
where $i$ and $j$ denote Cartesian components,
and $L$ is an appropriate length scale for the calculation 
of the gradient. If~we assume that the electric 
field varies linearly with the coordinates, then
$L$ would be the length scale over which 
the electric field ramps up from zero to its maximum value.
The magnitude $F_{\rm SO}$ of the spin-orbit force can, thus, be estimated as
\begin{equation}
F_{\rm SO} = 
\frac{\hbar |e| |\vec E| \sqrt{ g \, h }
\, \sin\theta} {2 \sqrt{2} L \, m_{e^+} c^2} \,.
\end{equation}
The gravitational force on the antihydrogen atom is 
$F_g = m_{\HHbar} \, g$, and~the ratio is
\begin{equation}
\frac{F_{\rm SO}}{F_g} = 
\frac{\hbar |e| |\vec E| \sqrt{ h }
\, \sin\theta} {2 \sqrt{2} \sqrt{g} \, L \, m_{e^+} \, m_{\HHbar} \, c^2} 
= 1.39 \times 10^{-14} \, 
\frac{ |\vec E|_{\rm SI} \, \sqrt{h_{\rm SI}} \, \sin\theta}{L_{\rm SI}} \,,
\end{equation}
where $|\vec E|_{\rm SI}$, $h_{\rm SI}$, and $L_{\rm SI}$
and the reduced quantities corresponding to the 
electric field, the~fall height $h$, and~the characteristic 
length scale $L$, expressed in SI mksA units, i.e.,
\begin{equation}
|\vec E|_{\rm SI} = |E| \, \frac{\rm m}{\rm V} \,,
\qquad 
h_{\rm SI} = \frac{h}{\rm m} \,,
\qquad 
L_{\rm SI} = \frac{L}{\rm m} \,.
\end{equation}
For $|\vec E|_{\rm SI} \sim 10^{6}$, 
$h_{\rm SI} \sim L_{\rm SI} \sim 1$, 
and $\theta \sim 1^\circ$, one has 
$F_{\rm SO}/F_g \sim 10^{-10}$.
Thus, the spin-orbit force can be neglected for our 
proposed experiment under realistic assumptions.
Intuitively, we could have guessed
this result on the basis of the fact that the entire
motion is fully nonrelativistic, and~the spin-orbit 
coupling term is a part of the Foldy--Wouthuysen 
transformed relativistic Dirac 
Hamiltonian~\cite{BjDr1964,ItZu1980}, 
which becomes significant only for relativistic systems.
In particular, the~spin coupling 
is highly suppressed in comparison to the direct coupling 
of the external electric field to any residual 
charge of antihydrogen, which would otherwise result from 
a putative charge asymmetry of~antimatter.

\end{document}